\documentstyle{mn}

\newif\ifAMStwofonts

\def\ltsima{$\; \buildrel < \over \sim \;$}
\def\lsim{\lower.5ex\hbox{\ltsima}}
\def\gtsima{$\; \buildrel > \over \sim \;$}
\def\gsim{\lower.5ex\hbox{\gtsima}}
\def\approxlt{\mathrel{\spose{\lower 3pt\hbox{$\sim$}}
        \raise 2.0pt\hbox{$<$}}}
\def\approxgt{\mathrel{\spose{\lower 3pt\hbox{$\sim$}}
        \raise 2.0pt\hbox{$>$}}}
\newcommand{\be}{\begin{equation}}
\newcommand{\en}{\end{equation}}
\def\deg {^\circ}

\def\mdot {\dot M}

\def\msole{~M_{\odot}}

\def\aa #1 #2 {A\&A {#1} #2}
\def\aas #1 #2 {A\&AS {#1} #2}
\def\araa #1 #2 {ARA\&A {#1} #2}
\def\mnras #1 #2 {MNRAS {#1} #2}
\def\apj #1 #2 {ApJ {#1} #2}
\def\apjs #1 #2 {ApJS {#1} #2}
\def\apjl #1 #2 {ApJ {#1} #2}
\def\aj #1 #2 {AJ {#1} #2}
\def\nat #1 #2 {Nature {#1} #2}
\def\pasj #1 #2 {PASJ {#1} #2}
\def\pasp #1 #2 {PASP {#1} #2}

\ifoldfss
  \ifCUPmtlplainloaded \else
    \NewTextAlphabet{textbfit} {cmbxti10} {}
    \NewTextAlphabet{textbfss} {cmssbx10} {}
    \NewMathAlphabet{mathbfit} {cmbxti10} {} 
    \NewMathAlphabet{mathbfss} {cmssbx10} {} 
  \fi
  \ifAMStwofonts
    \ifCUPmtlplainloaded \else
      \NewSymbolFont{upmath} {eurm10}
      \NewSymbolFont{AMSa} {msam10}
      \NewMathSymbol{\upi}     {0}{upmath}{19}
      \NewMathSymbol{\umu}     {0}{upmath}{16}
      \NewMathSymbol{\upartial}{0}{upmath}{40}
      \NewMathSymbol{\leqslant}{3}{AMSa}{36}
      \NewMathSymbol{\geqslant}{3}{AMSa}{3E}

    \fi
  \fi
\fi 

\ifnfssone
  \newmathalphabet{\mathit}
  \addtoversion{normal}{\mathit}{cmr}{m}{it}
  \addtoversion{bold}{\mathit}{cmr}{bx}{it}
  \newmathalphabet{\mathbfit} 
  \addtoversion{normal}{\mathbfit}{cmr}{bx}{it}
  \addtoversion{bold}{\mathbfit}{cmr}{bx}{it}
  \newmathalphabet{\mathbfss} 
  \addtoversion{normal}{\mathbfss}{cmss}{bx}{n}
  \addtoversion{bold}{\mathbfss}{cmss}{bx}{n}
  \ifAMStwofonts
    \ifCUPmtlplainloaded \else
      \UseAMStwoboldmath
      \makeatletter
      \new@mathgroup\upmath@group
      \define@mathgroup\mv@normal\upmath@group{eur}{m}{n}
      \define@mathgroup\mv@bold\upmath@group{eur}{b}{n}
      \edef\UPM{\hexnumber\upmath@group}
      \new@mathgroup\amsa@group
      \define@mathgroup\mv@normal\amsa@group{msa}{m}{n}
      \define@mathgroup\mv@bold\amsa@group{msa}{m}{n}
      \edef\AMSa{\hexnumber\amsa@group}
      \makeatother
      \mathchardef\upi="0\UPM19
      \mathchardef\umu="0\UPM16
      \mathchardef\upartial="0\UPM40
      \mathchardef\leqslant="3\AMSa36
      \mathchardef\geqslant="3\AMSa3E
    \fi
  \fi
\fi 

\ifnfsstwo
  \DeclareMathAlphabet{\mathbfit}{OT1}{cmr}{bx}{it}
  \SetMathAlphabet\mathbfit{bold}{OT1}{cmr}{bx}{it}
  \DeclareMathAlphabet{\mathbfss}{OT1}{cmss}{bx}{n}
  \SetMathAlphabet\mathbfss{bold}{OT1}{cmss}{bx}{n}
  \ifAMStwofonts
    \ifCUPmtlplainloaded \else
      \DeclareSymbolFont{UPM}{U}{eur}{m}{n}
      \SetSymbolFont{UPM}{bold}{U}{eur}{b}{n}
      \DeclareSymbolFont{AMSa}{U}{msa}{m}{n}
      \DeclareMathSymbol{\upi}{0}{UPM}{"19}
      \DeclareMathSymbol{\umu}{0}{UPM}{"16}
      \DeclareMathSymbol{\upartial}{0}{UPM}{"40}
      \DeclareMathSymbol{\leqslant}{3}{AMSa}{"36}
      \DeclareMathSymbol{\geqslant}{3}{AMSa}{"3E}
    \fi
  \fi
\fi 

\ifCUPmtlplainloaded \else
  \ifAMStwofonts \else 
    \def\upi{\pi}
    \def\umu{\mu}
    \def\upartial{\partial}
  \fi
\fi

\input psfig.tex

\title{New limits on the orbital parameters of 1E~1048.1--5937 and  1E~2259+586
from RossiXTE observations}

\author[S. Mereghetti et al.]
  {S.~Mereghetti,$^1$
  G.L.~Israel,$^{2,*}$
  and L.~Stella.$^{2,*}$\\
  $^1$Istituto di Fisica Cosmica del C.N.R., Via Bassini 15, I-20133 Milano,
Italy;  sandro@ifctr.mi.cnr.it \\
  $^2$Osservatorio Astronomico di Roma, Via dell'Osservatorio 2,
I-00040 Monteporzio Catone (Roma), Italy; \\ e-mail: (gianluca/stella)@coma.mporzio.astro.it\\
  $^*$Affiliated to ICRA.}

\date{Accepted 1997 October 23.
      Received 1997 September 25}

\pagerange{\pageref{firstpage}--\pageref{lastpage}}
\pubyear{1997}

\begin{document}

\maketitle

\label{firstpage}

\begin{abstract}
 
We report on two RossiXTE observations of the anomalous X--ray pulsars
1E~1048.1--5937 and 1E~2259+586. Both sources have continued their
almost constant spin-down during 1995/96. 
We carried out a search for orbital Doppler shifts, 
in their observed spin frequencies, deriving stringent limits on the
projected semi-axis. 
Unless these systems have unlikely small inclinations, main sequence
companions can be excluded. If   1E~1048.1--5937 and 1E~2259+586
are indeed binary systems, their companion stars must be either 
white dwarfs, or 
helium-burning stars with M $\lsim 0.8\msole$,  possibly underfilling their
Roche lobe.  

\end{abstract}


\section{Introduction} 

The two sources considered in this article belong to 
a small group of X--ray pulsars, with periods in the $\sim 5-10$~s 
range, that have recently attracted 
much interest owing to their peculiar properties
 (Mereghetti \& Stella 1995). In particular, 
the lack of bright optical counterparts implies that these 
neutron stars are not accreting from massive companions, 
contrary to the great majority of known X--ray pulsars.  Other 
characteristics that distinguish them from the more 
common pulsars in High Mass X--ray Binaries (HMXRBs) include 
very soft spectra,  X--ray luminosities of the order of
$10^{35}-10^{36}$~erg~s$^{-1}$, little long term variability, 
and relatively stable 
spin period evolution (see Stella et al. 1997 for a recent 
review). In the following we will refer to these systems as 
Anomalous X--ray Pulsars (AXP). 

The simplest interpretation for the AXP is that of Low 
Mass X--ray Binaries (LMXRBs) characterized by lower 
luminosity and higher magnetic field ($B\sim10^{11}$ G) than the 
classical, non-pulsating LMXRBs (Mereghetti \& Stella 1995). In 
this scenario, the observed period distribution, significantly 
different from that of HMXRB pulsars (which spans from 69 ms 
to 25 min), can be explained assuming that the  neutron 
stars in AXP are rotating at (or very close to) their equilibrium 
periods. 

However, the absence of orbital motion signatures, such as 
periodic delays of the pulse arrival times or periodic flux 
modulations, led also to non standard interpretations based on 
single stars (e.g. Paczy\'nski 1990; Thompson \& Duncan 1993; 
Corbet et al. 1995; Heyl \& Hernquist 1997). 
Recently it has been proposed that the AXP 
are the descendant of Thorne-Zytkow objects and consist of 
isolated neutron stars fed by a residual accretion disk (van 
Paradijs et al. 1995; Ghosh et al. 1997). 

Here we report the results of  RossiXTE observations of two AXP:
1E~1048.1--5937, serendipitously discovered 
with the Einstein Satellite  during observations of the Carina nebula
(Seward, Charles \& Smale 1986),
and 1E~2259+586, located at the center of the 
radio/X--ray supernova remnant G109.1-1.0 (Fahlman \& Gregory 
1981).  Since both sources have not been optically identified so 
far (Mereghetti, Caraveo \& Bignami 1992; Coe \& Jones 1992), 
X--ray observations are the  only available tool to assess their 
binary nature. The main objective of our observations was a 
search for orbital motions  through 
the detection of Doppler delays in the pulse arrival times. 
Though such a  search gave   negative results, the 
newly derived upper limits provide strong constraints on the 
possible companion stars for these two AXP.

\section[]{Search for Orbital Motion}

The observations of 1E~1048.1--5937 and 1E~2259+586 were performed with the 
RossiXTE satellite (Bradt et al. 1993) from 1996 July 29 22:44 
UT to 31 11:30 UT   (net exposure $\sim$86~ks) and from 1996 
September 29 15:40 UT to 30 3:05 UT  ($\sim$77~ks), 
respectively. The results presented here are based on data 
collected with the Proportional Counter Array (PCA, Jahoda et al. 
1996). The PCA instrument consists of an array of 5 
proportional counters operating in the 2--60 keV energy range, 
with a total effective area of approximately 7000~cm$^{2}$  and a 
field of view, defined by passive collimators, of $\sim1\deg$   
FWHM. Most of our analysis is based on data collected by the 
on board electronics in the so called "Good Xenon" operating 
mode. This provides the time of arrival and the pulse height 
of each count.  In order to reduce the 
background, only the counts detected in the first Xenon layer of 
each counter were used.  

For each source we determined the spin period 
with a   standard folding technique, after converting  
to the Solar System barycenter
the times of arrival of the counts.
We obtained the values  $P= 6.449769\pm0.000004$~s for 1E~1048.1--5937
and  $P= 6.978912\pm0.000003$~s for  1E~2259+586. 
The corresponding light curves    are similar to those previously 
observed from these 
sources with other satellites.
Our period values show that both sources have continued
their secular spin-down during 1995/96 (see
Fig. 1).  The spin period of 1E~2259+586 is smaller than
predicted from a linear extrapolation of the last measurements 
obtained with ROSAT and ASCA. 
The opposite situation occurs for  1E~1048.1--5937, the 
spin-down rate of which has further increased.

\begin{figure}
\centerline{\psfig{figure=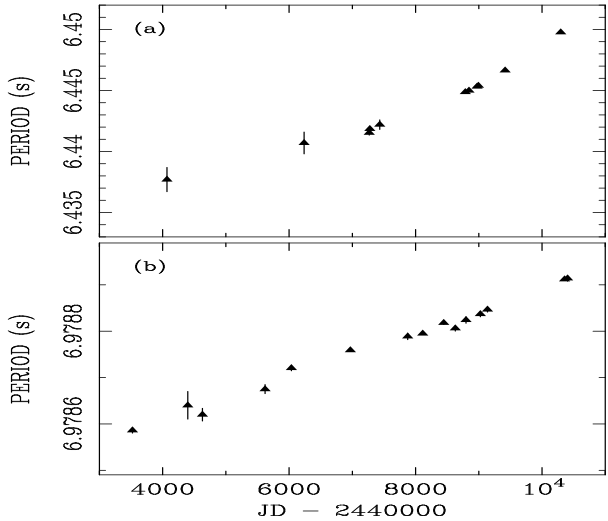,width=8cm,height=6.5cm} }
 \caption{Long term   period evolution of  1E~1048.1--5937 (a) 
and  1E~2259+586 (b).}
\end{figure}

\begin{figure*}
\centerline{\psfig{figure=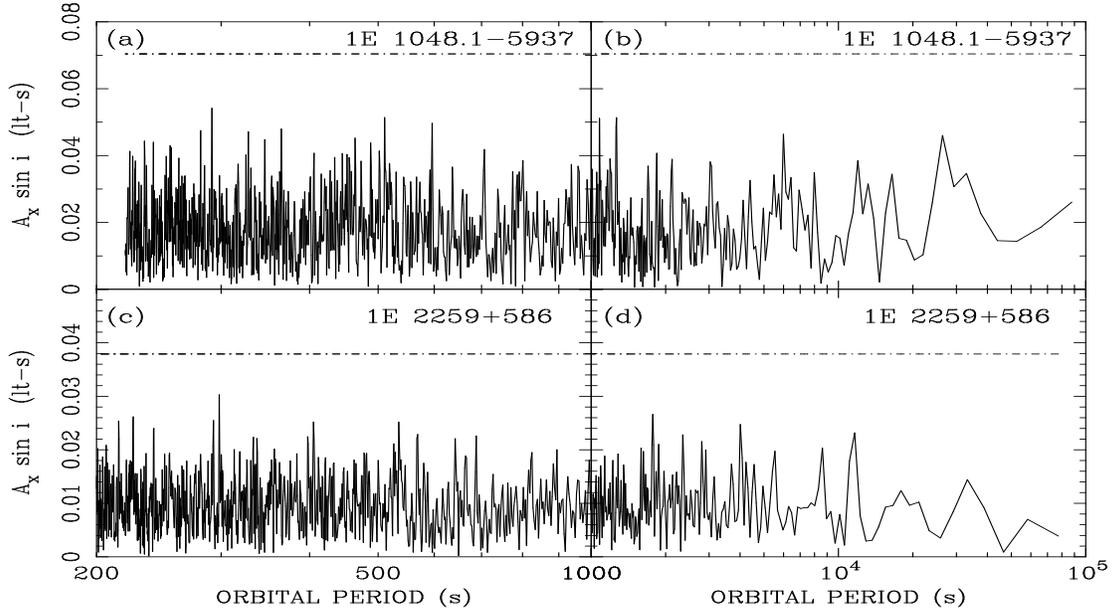,width=8cm,height=7.5cm} }
  \caption{Results of a search for orbital Doppler delays for  1E~1048.1--5937 (a,b) 
and  1E~2259+586 (c,d). The dashed lines indicate the threshold values on $a_x\sin i$
corresponding to detections at the 99\% confidence level. For each source 1200 trial
orbital periods between $\sim 200$~s and $\sim 1$~d have been searched.}
\end{figure*}

In the search for orbital Doppler modulations 
in 1E~1048.1--5937, we considered 1200 trial orbital periods between $\sim$200~s 
and $\sim1.3 \times 10^{5}$~s oversampling the Fourier resolution
dictated by the time span of the observation by a factor 2. 
For each trial period, we computed 8 light curves, folded at the 
spin period value, corresponding to different intervals of the 
orbital phase. Particular care was devoted to clean the data 
from intervals with anomalously high background and to 
properly take into account the effective exposure of every 
phase bin. The resulting light curves were cross-correlated with 
the template one, obtained using all the data, and the peaks in 
the  cross-correlation curves were then fitted with a Gaussian. 
The Gaussian centroids  provide a measurement of possible time 
delays. To search for the sinusoidal modulation expected in the 
8 delays in the case of a circular orbit, we computed their 
Fourier amplitudes. These are shown in Fig. 2a and 2b as a 
function of the trial orbital period. As expected, the squared 
Fourier amplitudes are distributed according to a  $\chi^2$ with 2 
degrees of freedom. The presence of an orbital modulation would be revealed 
by a peak exceeding the threshold indicated by the dashed line 
(corresponding to a 99\% confidence level).  No significant peaks 
were  found and, following van der Klis (1989),  we derive an 
upper limit of  $a_x\sin i < 0.06$~lt-s   (99  \% c.l.). 

The same procedure was applied to 1E~2259+586, for 1200 periods 
between $\sim194$~s and $\sim1.2 \times 10^{5}$~s 
(see Fig. 2c and 2d).  Also in this case no 
significant orbital modulations were found, with an upper limit 
of   $a_x\sin i< 0.03$~lt-s.

Finally, we  searched for a periodic modulation in the  source light curves
by means of a Fourier analysis,
 without finding any statistically
significant periodicity.
Using the method of Groth (1975) and  Vaughan et al. (1994), modified as 
described in Israel \& Stella (1996),
we derived the following  3 $\sigma$ upper limits on the flux pulsed
fraction for the two sources: 
2\% for periods shorter than 1000 s,
$\sim$6\% for periods between 1000 s and one hour,
and from $\sim$6\%  to  $\sim$20\% for increasing period
values up to $\sim$5 hours.
For comparison,  
4U~1820--303, with an orbital period of  11~min,
has a peak to peak modulation of 3\% 
(Stella, Priedhorsky \& White 1987).

\section{Discussion}

Previous observations of 1E~1048.1--5937 with the GINGA satellite (Corbet \& 
Day 1990) yielded an upper limit of $a_x\sin i< 0.6$~lt-s  for orbital 
periods between 100 and 1100 s. The source was reobserved 
with ROSAT in 1992/93 (Mereghetti 1995), with ASCA in 1994 
(Corbet \& Mihara 1996), and recently with BeppoSAX (Oosterbroek  
et al. 1997). These observations confirmed the long term 
spin-down at a rate of $\sim9\times10^{-4}$~s~yr$^{-1}$, but did 
not allow more 
sensitive searches for orbital modulations. Our RossiXTE 
observation has provided an upper limit on $a_x\sin i$ one order of 
magnitude smaller than the previous value and applicable to a wider range 
of possible orbital periods. Also in the case of 1E~2259+586 our 
limits are stronger than the previous value of  $a_x\sin i< 0.08$~lt-s for 
$1000 < P_{orb} < 10000$~s derived with GINGA (Koyama et al. 
1989). 

In Fig. 3 the new limits on the orbital parameters are plotted 
for representative values of the unknown inclination angle. Our 
results significantly reduce the allowed parameter space with 
respect to the previous limits, setting strong  constraints on the 
nature of the possible companion stars of these pulsars.

The evolution of LMXRBs with orbital periods as short as 
those considered here ($\lsim$few hours), 
is mainly driven by angular momentum losses due  
to gravitational radiation. In this case, and with the 
usual assumption  of conservative mass transfer,  
a well defined relation between the mass $M_{c}$ of a 
companion filling the Roche-lobe and the orbital period has to 
be satisfied (see, e.g., Verbunt \& van den Heuvel 1995). This 
relation depends on the nature of the companion star and has 
been reported on Fig. 3 for three different cases corresponding to
a main sequence star, a helium burning star, 
and a hydrogen depleted, fully degenerate white dwarf.

\begin{figure*}
\centerline{\psfig{figure=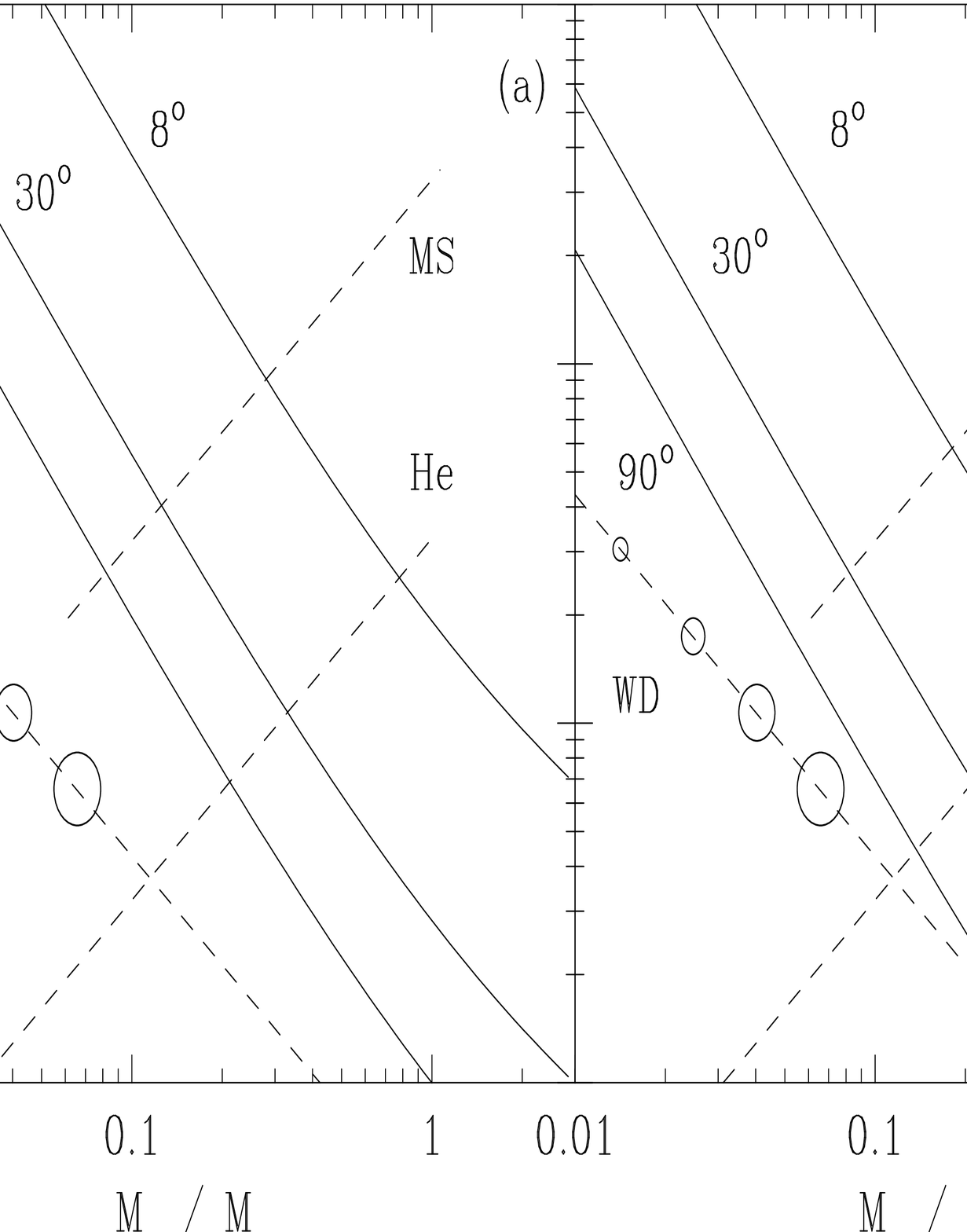,width=8cm,height=5.5cm} }
  \caption{ Constraints on the companion mass and orbital period 
for 1E~1048.1--5937 (a) and  1E~2259+586 (b). 
The three lines refer to orbit inclinations of $90\deg$, $30\deg$ and $8\deg$.
A neutron star mass of 1.4 $\msole$ has been assumed. 
Regions above the full lines are excluded by the limits on $a_x\sin i$.
The dashed lines 
indicate the orbital period at which stars of different type  fill their Roche lobe
for a given mass (see Discussion).
The four circles of increasing size correspond to values of 
$\mdot =  10^{-11}, 10^{-10}, 10^{-9}, 10^{-8} ~\msole$~year$^{-1}$.
}
\end{figure*}

\subsection{A main sequence companion}

The maximum mass of a hydrogen main sequence companion 
filling the Roche lobe is limited to $\sim0.1 \msole$ for 
inclinations greater than 30$\deg$. Only in the case of a very 
small inclination, larger masses would be compatible with our 
limits on $a_x\sin i$. For example, if $i=8\deg$   the companion
of  1E~1048.1--5937
could be as massive as $0.3 \msole$ ($0.2 \msole$ for 1E~2259+586).
Note, however, that the chance probability of 
observing a binary system with an inclination smaller than this 
is only 1\%. 

The faintness of the possible optical counterparts of 1E~2259+586
(Coe  \& Jones 1992) only allows companions of late-K or M 
spectral type.  Since K stars have masses of the order of $0.5-1 
\msole$, the presence of such a star can now be considered very 
unlikely (both in 1E~1048.1--5937 and 1E~2259+586),
leaving only a very late-M dwarf as a viable possibility.  

\subsection{A helium burning companion}

Higher masses (up to $\sim0.8 \msole$) are allowed in the case of
helium-burning companions filling their Roche lobe. 


The measured X--ray fluxes are in the range 
$( 0.6 - 3 )\times 10^{-11}$~erg~cm$^{-2}$~s$^{-1}$ for 1E~1048.1--5937
(Seward et al. 1986; Corbet \& Mihara 1996) and 
$(2-5)\times 10^{-11}$~erg~cm$^{-2}$~s$^{-1}$ for 1E~2259+586 
(Iwasawa, Koyama \& Halpern 1992; Parmar et al. 
1997). 
Though the distances  are uncertain, they are quite well constrained by
the size of the Galaxy. In fact both sources are in the galactic plane, and, 
taking 20 kpc as a reasonable upper limit to their distance,
we obtain luminosities  
$L_{x} = 4\times 10^{35}~(F/10^{-11}~{\rm erg~cm^{-2}~s^{-1}})~(d/20~{\rm 
kpc}){^2}$~erg~s$^{-1}$.

The corresponding accretion rate  
($\mdot = 4\times 10^{-11}~(F/10^{-11}~{\rm erg~cm^{-2}~s^{-1}})~(d/20~{\rm 
kpc})^{2}~\msole$~yr$^{-1}$) is much lower than the value
of $\sim3\times10^{-8}~\msole$~yr$^{-1}$
expected for Roche lobe overflow accretion
from a  $\sim0.8 \msole$ helium star (Savonije, Kool \& van den Heuvel 1986). 
Even higher  accretion rates would be expected  
for lower masses and correspondingly shorter orbital periods of 
a few minutes.  

Thus a more likely possibility is that the 
companion underfills the Roche lobe and accretion is via stellar 
wind. Such a case has been considered by Angelini et al. (1995) 
for 4U~1626--67, another source included in the group of AXP 
(Mereghetti \& Stella 1995) despite having some remarkable 
difference with respect to the other objects. In particular, 4U 
1626--67 has an optical identification and a well established 
orbital period of 42 min (Middleditch et al. 1981; Chakrabarty 
1997). Though  this led other authors to exclude it from the 
sample of AXP, it has been proposed (Ghosh et al. 1997)  that 
4U~1626--67, as well as other short period binaries such as Cyg 
X-3 (van Kerkwijk et al. 1992) and HD 49798 (Israel et al. 1997), 
might result from an evolutionary scenario 
similar to that of AXP, involving a common envelope phase of a 
progenitor HMXRB.  

\subsection{A white dwarf companion}

It can be seen in Fig. 3 that, 
due to the different $M_c - P_{orb}$ dependence in the case of accretion from a white dwarf,
the limits on $a_x\sin i$  do not constrain the companion mass.   
A different way to derive some information on the  companion   is to 
consider  the mass accretion rate expected 
for different masses and compositions of the white dwarf.

For conservative accretion through Roche lobe overflow, the 
expected accretion rate depends strongly on the mass of the white 
dwarf companion. Different values of $\mdot$,
ranging from $10^{-11}$ to $10^{-8} ~\msole$~year$^{-1}$, have been
indicated in Fig. 3 on the white dwarf $M_c - P_{orb}$ relation.
Values compatible with the observed 
luminosity limits can easily be obtained (Savonije et al. 1986).  
For example, if we use for 1E~2259+586 the distance estimates for the 
G109.1-1.0 supernova remnant (from 3.6 to 5.6 kpc,   
Hughes et al. 1984) we derive  $\mdot\sim3\times 10^{-11}~\msole$~yr$^{-1}$. 
Such 
a value requires a white dwarf mass of $\sim0.02~\msole$, and the 
corresponding orbital period would be of the order of 30 min. 
We note that a white dwarf was discovered among the 
possible optical counterparts of 1E~2259+586  (star E of Davies \& Coe 1991). 
However, if this star has colours and 
luminosity of a typical white dwarf, it must be at a distance
of $\lsim$1 kpc, incompatible with that of the SNR. 
Thus, if star E, the only object with unusual colours in the 
field (Coe \& Jones 1992),  is the counterpart   one has to invoke a
chance coincidence between 1E~2259+586 and  G109.1-1.0 
(the alternative possibility that     
star E be  more luminous than a white dwarf runs into problems
since, at a distance greater than 3--4 kpc
it should also appear more reddened).

\section{Conclusions}
 
The large collecting area of the RossiXTE PCA instruments has allowed a very sensitive
search for orbital light propagation delays
in the pulsations of  1E~1048.1--5937 and  1E~2259+586. 
The   upper limits on  $a_x\sin i$ derived from these observations 
have greatly reduced the range of possible companion stars and orbital parameters
for these sources.

Main sequence companions are now virtually excluded, leaving only either
helium-burning stars with M $\lsim 0.8\msole$ or white dwarfs as possible
mass donors. Moreover, the low luminosity of these systems 
excludes orbital periods shorter than $\sim$ 1000 s in the white dwarf case,
while in the helium star case it favours mass transfer through a stellar wind.

Though our result supports models for AXP based on isolated neutron stars,
a further sensitivity improvement  is required to 
completely  rule out binary model (barring the case $i\sim0\deg$). 
 
At least in the case of  1E~2259+586, a longer observation with a large collecting area 
instrument should  reveal the presence of an orbital motion or definitely exclude
the possibility of a white dwarf companion, independently from its mass.

\section*{Acknowledgments}
We thank S. Campana for helpful discussions.

\label{lastpage}

\end{document}